\documentclass[preprint]{revtex4}
\usepackage{graphicx}

\begin{document}

\title{Dynamics of an Exciton-phonon Triangle under Photoirradiation}


\author{Noriyuki Aoyagi}
\affiliation{Graduate School of Regional Development and Creativity, 
Utsunomiya University, Utsunomiya, Tochigi 321-8585, Japan}
\author{Hiroaki Matsueda}
\affiliation{Department of Applied Physics, Graduate School of Engineering, 
Tohoku University, Sendai 980-8579, Japan}
\author{Kunio Ishida}
\email{ishd\_kn@cc.utsunomiya-u.ac.jp}
\affiliation{School of Engineering and Center for Optical Research and Education, Utsunomiya University,
Utsunomiya, Tochigi 321-8585, Japan}

\begin{abstract}
Herein, the dynamics of excitons coupled with optical phonons in a triangular system is numerically studied.
By representing the excitons by quasi-spin states, the similarity between the chiral spin states and the exciton chiral states is discussed.
In particular, the optical control of excitons is discussed, where photoirradiation causes the switching of the exciton states on the ultrafast time scale by Raman scattering .

A phase diagram is obtained based on the ground-state properties of the system determined by the magnitudes of the exciton-phonon interactions and exciton transfer energy. 
By varying the frequency and/or intensity of light, a transition between exciton-phonon composite states is induced, which suggests the possibility of the coherent control of the chiral properties of excitonic systems via phonon excitation.
\end{abstract}

\maketitle

\section{Introduction}
\label{intro}

The recent progress of ultrashort pulse laser technology has rendered observing the time evolution of quantum-mechanical states in the coherent regime possible\cite{measure2,measure3,measure4,measure5}, which renders studying the light control of quantum many-body states possible particularly regarding the transient properties of condensed matter.

The generation and the storage of quantum entanglement between remote systems has attracted our attention, where nonlocal correlation between qubits plays an essential role.
For example, recent experimental studies have reported the methods of light-mediated entanglement generation between non-interacting systems\cite{lee,borjans}.
The mechanism of photoinduced entanglement generation was studied by the present authors, who found that the phonon degrees of freedom are suitable for quantum entanglement storage between remote systems\cite{mejpsj}.
By extending the model of remote material systems, we studied the entanglement dynamics of electron-phonon systems wherein the interplay of electron exchange, electron-phonon, and dipole interactions plays an important role in the time-evolution of the spin-phonon states\cite{mefd}.
More precisely, the entanglement between them generates non-classical states of spin-phonon systems, e.g., the dynamical Jahn-Teller states\cite{JT}, which shows that the effect of the intersite-geometry between localized electronic systems should be clarified to understand the possible quantum-mechanical states obtained by coherent control methods.
These types of quantum-mechanical control are reminiscent of ``quantum materials'' realized as transient states of materials under an external field\cite{qm}, where profound knowledge of quantum many-body states is required to obtain a material design method in the transient regime.

In this paper, we investigate the dynamics of exciton-phonon systems in which three exciton sites are located in a triangle.
When material systems have triangular symmetry, the chiral properties of electrons play an important role in various physical properties.
In an antiferromagnetic spin-triangle, for example, spin-frustrated states appear as the ground state whose chiral degree of freedom shows particular magnetic properties.
In particular, the effects of the vibronic coupling in the $(E \times e)$ Jahn-Teller systems are understood in various aspects\cite{Exe1,Exe2}.

An exciton triangle has also attracted our attention in molecular aggregate\cite{mola} and the quantum phase transition in triple quantum dots has been discussed\cite{dots}.
This system shows frustration properties\cite{dots3,dots4,dots5,dots6,dots7} which are considered as a quantum gate\cite{dots2}. 
The Kondo effect in a similar system was also studied when the electronic correlation plays an important role\cite{dots8}. 
The electronic states in these systems are described by the chiral properties of the wavefunction, and the exciton-phonon entanglement caused by the aforementioned interactions also provides us with various quantum states.
Precisely, the chiral properties of an exciton triangle reflect on its optical properties, e.g., optical rotation, and thus the coherent control of excitonic properties will realize an optical device to switch the polarization of light. 
Hence, we aim to discuss their dynamical properties through numerical calculations to understand the coherent control method of exciton-phonon systems.

\section{Model and method}
\label{mome}

We study the quantum dynamics of triangular exciton-phonon systems under light irradiation.
We consider that the exciton is described by a two-level system for simplicity, which means that only a single exciton is excited at each site of the triangle.
In other words, we consider only weak-excitation cases.
The exciton states are described by two kets $|g\rangle$ and $|e\rangle$, where the former shows no excitons, and the latter shows a single exciton.
In this case, linear operations on these states are described by the Pauli matrices, where
\begin{eqnarray}
\sigma_z |g\rangle & = & -|g\rangle\\
\sigma_z |e\rangle & = & |e\rangle.
\end{eqnarray}
In this way, we derived a model of exciton-phonon systems of which the Hamiltonian is described by\cite{mefd,tc},
\begin{eqnarray}
  {\cal H}_0  &  =  & \sum_{j=1}^3 \left [ \omega  a_j^\dagger a_j + n_j \{\varepsilon + \nu(a_j^\dagger +a_j)  \} \right ] + J\sum_{\langle i,j \rangle} \vec{\sigma}_i \cdot \vec{\sigma}_j \nonumber \\
  & = & \sum_{j=1}^3 \left [ \omega  a_j^\dagger a_j + n_j \{\varepsilon + \nu(a_j^\dagger +a_j)  \} \right ] + J\sum_{\langle i,j \rangle} \{ 2(\sigma^+_i \sigma^-_j + \sigma^-_i \sigma^+_j)+  \sigma^z_i \sigma^z_j \},
\label{ham}
\end{eqnarray}
where $\sigma_i^\alpha\ (\alpha = x,y,z)$ denotes the Pauli matrix corresponding to the exciton at site $i$, and $\sigma_i^\pm = (\sigma_i^x \pm i\sigma_i^y)/2$.
Thus, $\sigma^+_i \sigma^-_j$ shows the nearest-neighbor hopping of the exciton from site $j$ to $i$, i.e., the transfer energy of the exciton is $2J$.
$n_j = (\sigma_j^z\ + 1)/2$ also denotes the exciton population at site $j$, and
$\sigma^z_i \sigma^z_j$ is the dipole interaction between excitons in different sites.
We fix the exciton tranfer energy twice as large as the dipole interaction strength for clear discussion on the excitonic properties in the present study.  
In other words, we describe the electron-phonon systems by a spin-boson model in which a "quasi-spin" state represents a Frenkel exciton at each site, and the phonons are treated as the Einstein phonons.

\begin{figure}
  \centering
  \includegraphics[width=8cm,pagebox=cropbox,clip]{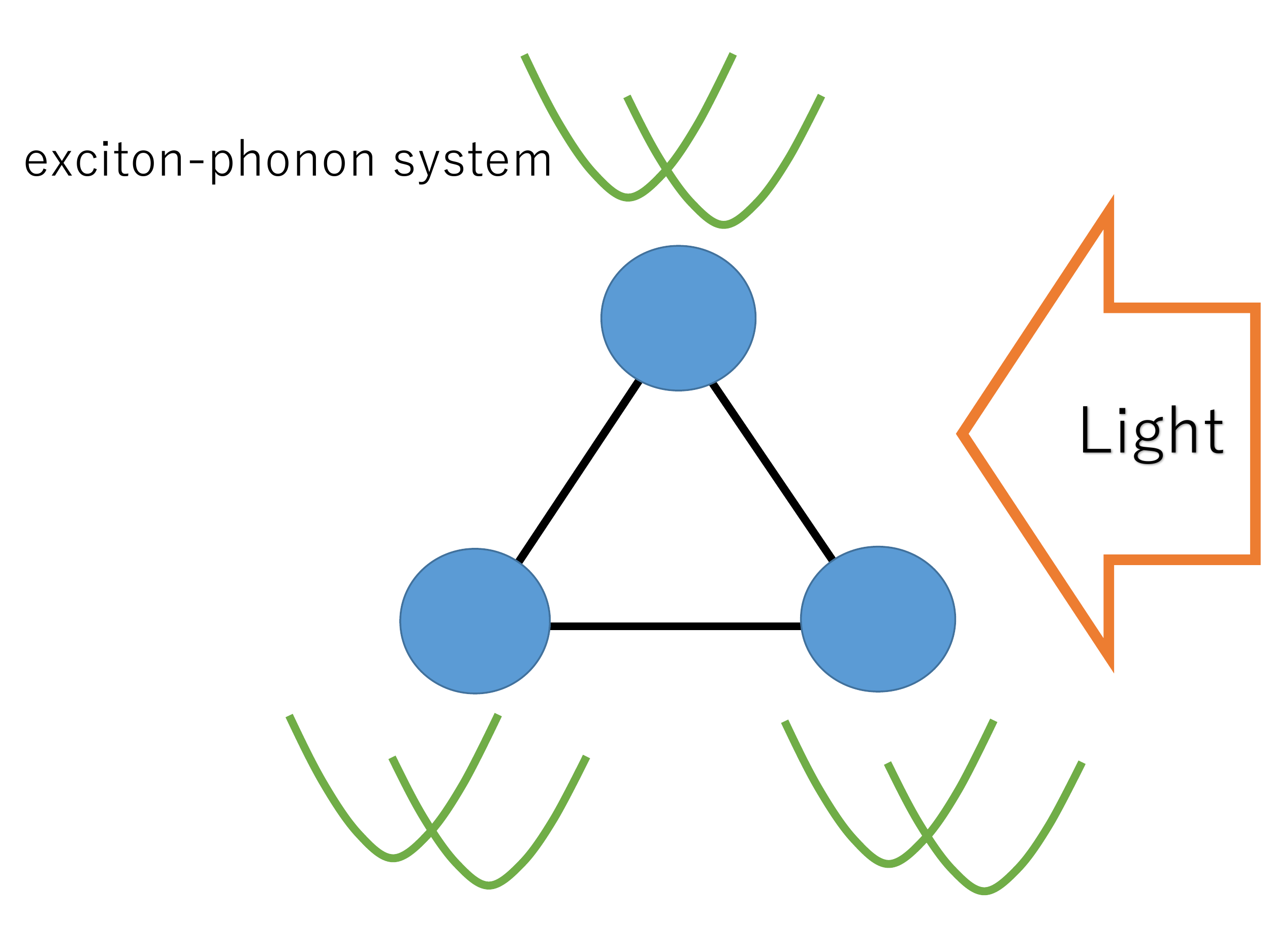}
  \caption{A schematic of the model described by Eq.\ (\ref{ham}).}
  \label{modelview}
\end{figure}

The values of the parameters are: $\omega = 1$ and $\varepsilon = 8$.
The other parameters are variable as discussed in the following sections.
When $\varepsilon = 1$eV, $\omega$ is $\sim 100$meV, which is comparable to the frequency of optical phonons in organic molecules or semiconductors.
Since the bandwidth of excitons in molecular aggregates is similar to or less than $\varepsilon$\cite{moldisp}, the value of $J$ is less than $\varepsilon/2$ in this paper.
As for the value of $\nu$, we assume that the average number of excited phonons $\nu^2$ is at most 4, which means $\nu < 2$.
A schematic of the model is shown in Figure \ref{modelview}.

As mentioned in Section \ref{intro}, the chirality of the quasi-spins is important to discuss the symmetry properties of  the system, since it describes the chiral properties of the dipole induced by optical excitation.
In this paper we discuss the chiral properties of the exciton state by
\begin{equation}
\chi = {1 \over 2\sqrt{3}} \vec{\sigma}_1 \cdot (\vec{\sigma}_2 \times \vec{\sigma}_3),
\label{chi}
\end{equation}
in analogy to the scalar spin chirality in spin systems.
As $\chi$ commutes with both $\sum_{\langle i,j \rangle} \vec{\sigma}_i \cdot \vec{\sigma}_j$ and $\sum_i \sigma_i^z$, $\chi$ is a good quantum number for $\nu=0$.
However, in the presence of electron-phonon interaction, $[{\cal H}_0, \chi] \neq 0$, and thus $\chi$ is not conserved any more like $s = |\sum_j \vec{\sigma}_j|^2$.
  
We assume that the triangular system in the present study does not have inversion symmetry owing to the structural properties of each site, for example, which is introduced by certain types of electronic interactions, e.g., the Dzyaloshinskii–Moriya (DM) interaction\cite{dzy,moriya}.
In fact, a recent study has shown that a DM-like interaction is present in ferroelectrics and antiferroelectrics\cite{dpdm}.
In the present study, we assume an infinitesimal DM interaction between dipoles which makes a state with finite value of $\chi$ the ground state of the system, although its effect on the dynamics of the system is ignored.

In the present study, we consider light irradiation which is described by the interaction Hamiltonian,
\begin{equation}
{\cal H}_i = \mu E_0 \cos \Omega t \sum_j\sigma_j^x ,
\end{equation}
where $\mu$ and $E_0$ denote the transition dopole moment for an exciton and the electric field of the light, respectively.
We do not apply the rotating wave approximation in the present study, since we vary the value of $\Omega$ and calculated the photoexcitation dynamics for both resonant and non-resonant cases.
We numerically solved the time-dependent Schr\"odinger equation,
\begin{equation}
i \frac{d}{dt}|\Psi(t) \rangle = ({\cal H}_0+{\cal H}_i)|\Psi(t)\rangle,
\end{equation}
to study photoinduced dynamics of the exciton-phonon system.
The wavefunction $|\Psi(t)\rangle$ can be expanded as
\begin{equation}
|\Psi(t) \rangle = \sum_{\sigma_1  \sigma_2 \sigma_3 n_1 n_2 n_3} p_{\sigma_1  \sigma_2 \sigma_3 n_1 n_2 n_3} |\sigma_1 \sigma_2 \sigma_3 n_1 n_2 n_3 \rangle,
\end{equation}
where $|\sigma_i n_i \rangle\ (\sigma_i = g, e,\ n_i = 0, 1, 2,...)$ denotes the exciton-phonon state at site $i$.
The coefficients $p_{\sigma_1  \sigma_2 \sigma_3 n_1 n_2 n_3}$ are calculated by the fourth-order Runge-Kutta method, for which the time step was set to $2^{-15}\pi \sim 9.6\times 10^{-5}$, and the wavefunction was initially set to the ground state of  Hamiltonian ${\cal H}_0$.

\section{Calculated results}

\subsection{Ground-state phase diagram on $J$-$\nu$ plane}

When $J$ is small, the electronic configuration of the exciton system is given by the exciton ground state $\propto |g g g \rangle$, because the exciton excitation from $|g \rangle$ to $|e \rangle$ costs energy $\varepsilon$.
As $J$ increases, the energy gain from the exciton tranfer overcomes the energy cost for the exciton excitation.
This competition between $\varepsilon$ and $J$ gives us qualitatively different ground state by varying $J$.
Precisely, an entangled exciton-phonon state becomes the ground state by increasing $J$, which we call a chiral state in this paper, as $\langle \chi \rangle$ is finite.
On the other hand, the electron-phonon interaction lowers the energy of an exciton owing to the lattice relaxation, and thus the transition between the exciton ground state and the chiral state takes place at a smaller value of $J$ than that for $\nu = 0$.

\begin{figure}
  \centering
  \includegraphics[width=8cm,pagebox=cropbox,clip]{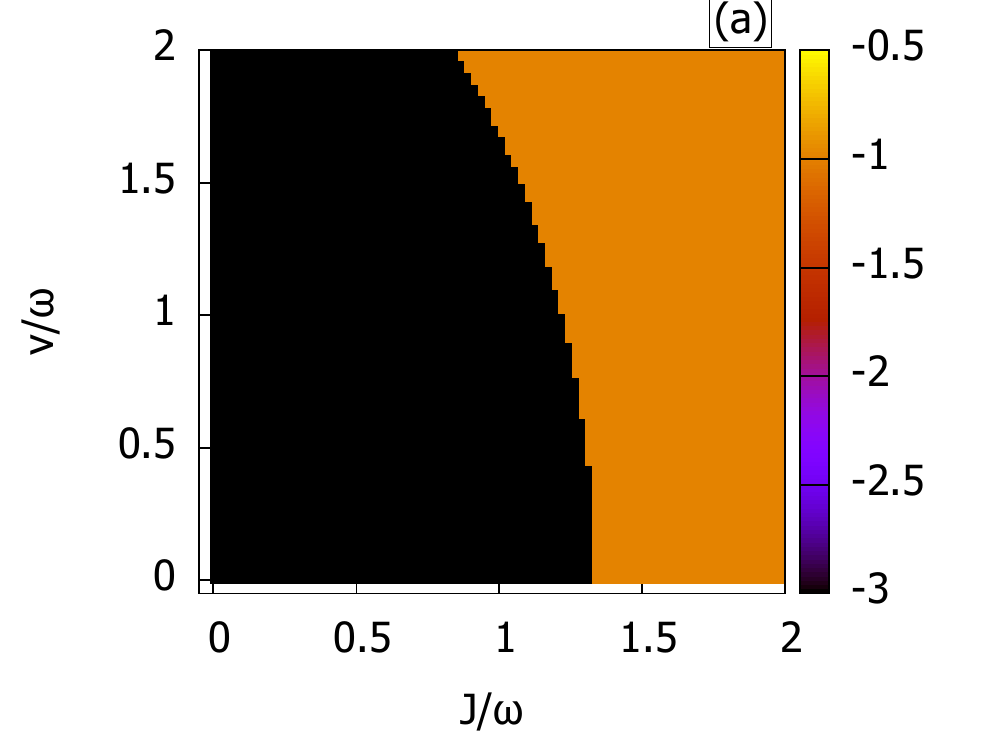}
  \includegraphics[width=8cm,pagebox=cropbox,clip]{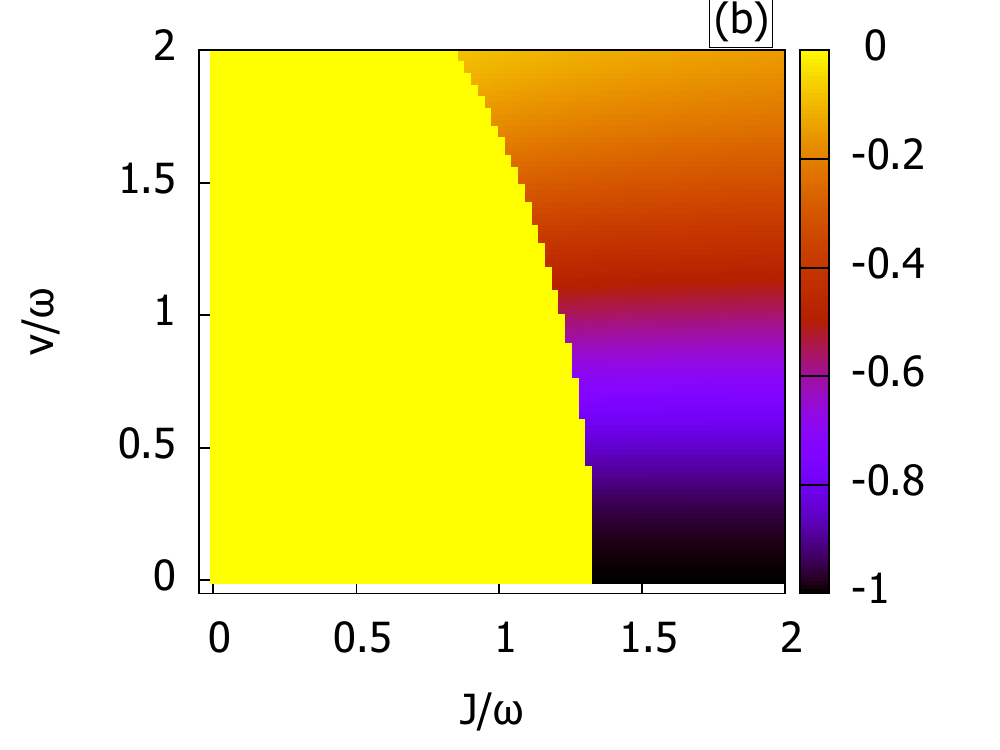}
  \caption{Phase diagram of the system characterized by the ground state property depicted by (a) $\langle \sum_j \sigma_z^j \rangle$ and (b) $\langle \chi \rangle$.}
  \label{phase}
\end{figure}

The above consideration suggests that we draw a phase diagram that characterizes the ground state of the material system.
For this purpose, we numerically diagonalized ${\cal H}_0$ by varying the values of $J$ and $\nu$, and calculated $N=\sum_i n_i$ and $\chi$.
Figure \ref{phase}-(a) shows the phase diagram on the $J$-$\nu$ plane, where the values of $N$ are depicted in colors.
Figure \ref{phase} shows the boundary between the two phases, i.e., phase 1 ($\sum_j \sigma_z^j=-3$) and phase 2 ($\sum_j \sigma_z^j=-1$).
Figure \ref{phase}-(b) also shows the phase diagram where the color map indicates the values of $\chi$.
$\chi$, which is always 0 for phase 1, varies in phase 2.
We note that maximum values of $J$ and $\nu$ are 2 in the present study, since the states with two or three excitons become a ground state for larger value of $\nu$. 
In this case, the excited states have less excitons than the ground state, which means that the role of $|g\rangle$ and $|e\rangle$ is reversed.
Hence, we do not discuss such "unphysical" situations in the present study.

The exciton-phonon state in phase 1 is $|\Psi \rangle = |g g g 0 0 0 \rangle$.
In this case, the total energy of the material system $E$ is given by $E=-3J$ and the excitons and phonons are decomposable.
Both the magnitude of the total quasi-spin $|\sum_n\vec{\sigma}|^2$ and $\chi$ are good quantum numbers in phase 1, as in isolated electronic systems.

As in the case of two-site systems\cite{mefd}, we discuss the ground-state properties in phase 2 by a variational wavefunction
\begin{equation}
  |\Psi_v  \rangle = {1 \over \sqrt{3}}(|g g e \beta  \beta \alpha \rangle + \theta|g e g \beta \alpha  \beta\rangle  + \theta^2 |e g g \alpha \beta \beta\rangle ),
  \label{vari}
\end{equation}
where $\theta = e^{2\pi i/3}$, and $|\alpha \rangle $ and $|\beta \rangle$ denote the coherent states of the phonon parametrized by $\alpha$ and $\beta$, respectively.
We note that $|\Psi_v \rangle$ is a good approximation for the ground state wavefunction in phase 2 particularly for small value of $\nu$ as shown in the supporting informations\cite{supp}.
However, it also helps us understand the entanglement properties of the ground state wavefunction qualitatively.

$\alpha$ and $\beta$ are determined by minimizing the energy expectation value for $|\Psi_v\rangle$, which is given by
\begin{equation}
E = \langle \Psi_v |{\cal H}_0|\Psi_v\rangle = \omega (|\alpha|^2 + 2|\beta|^2)-J(1+2e^{-|\alpha-\beta|^2})+\nu(\alpha + \alpha^\ast)+\varepsilon.
\end{equation}
Differentiating $E$ with respect to $\alpha$ and $\beta$, we obtained the following set of equations,
\begin{eqnarray}
\label{eqa}
 \omega \alpha +2J(\alpha-\beta)e^{-(\alpha-\beta)^2}+\nu &= & 0\\
 \omega \beta - J (\alpha-\beta)e^{-(\alpha-\beta)^2} & =  & 0,
\label{eqb}
\end{eqnarray}
where we consider that both $\alpha$ and $\beta$ are real numbers at the minimum point of $E$.
The relationship between $\alpha$ and $\beta$ is directly derived from the above equations as
\begin{equation}
\beta = -\frac{\alpha}{2}-\frac{\nu}{2\omega}.
\end{equation}

We note that $\alpha=-{\omega \over \nu}$ and $\beta=0 $ for $J=0$, and $\alpha=\beta$ if and only if $\nu=0$.
The numerically obtained wavefunction in phase 2 is well-approximated by Eq.\ (\ref{vari}) by taking the values of $\alpha$ and $\beta$ as the solutions of Eqs.\ (\ref{eqa}) and (\ref{eqb}).
Thus, the dynamical Jahn-Teller effect plays an important role in determining the exciton-phonon states in phase 2.

\begin{figure}
  \centering
  \includegraphics[width=8cm,pagebox=cropbox,clip]{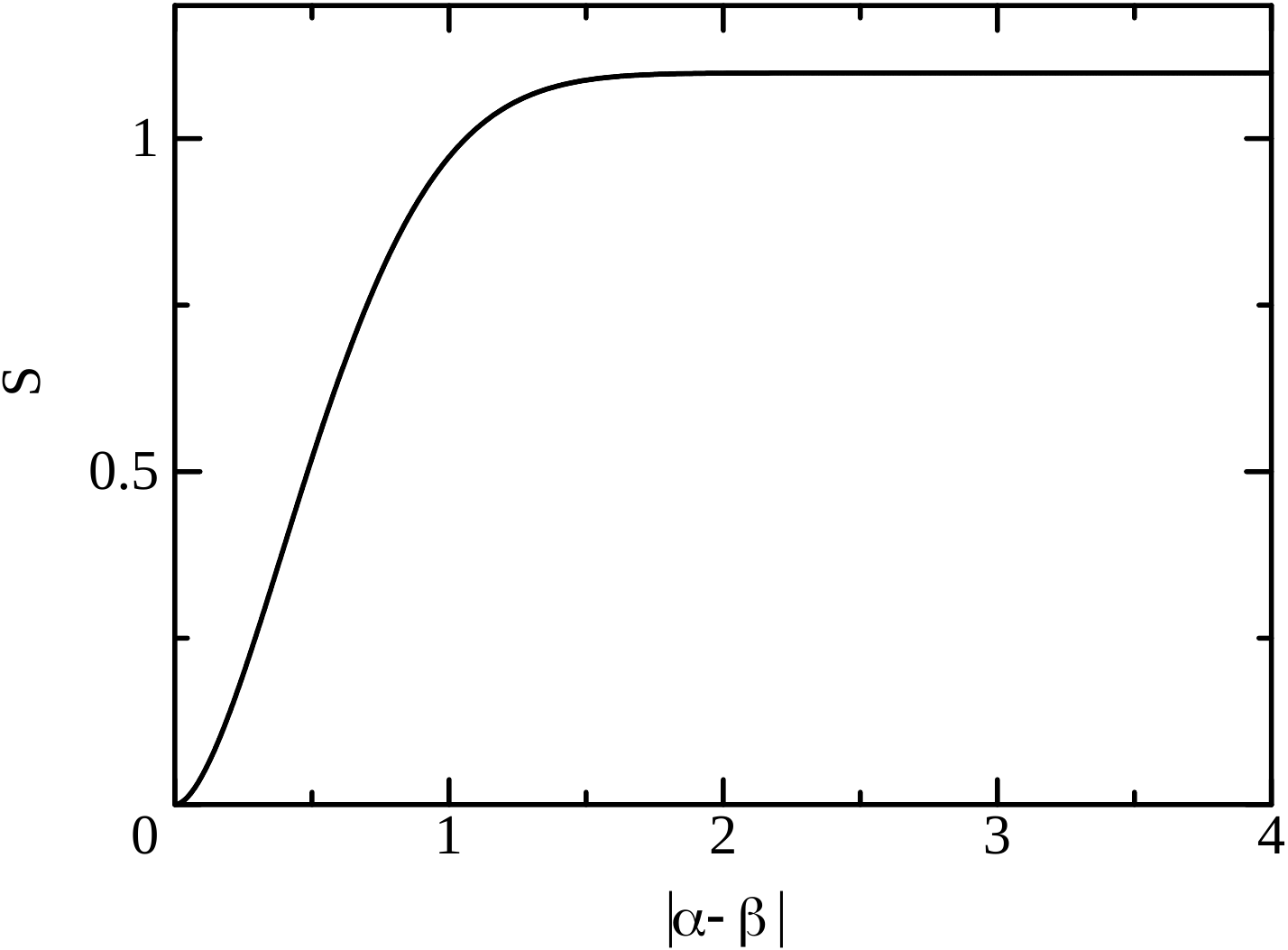}
  \caption{Entanglement entropy between excitons and phonons as a function of $|\alpha -\beta|$ obtained from the variational wavefunction $\Psi_v$.}
  \label{ee}
\end{figure}

In other words, the excitons and phonons are entangled with each other for $\alpha \neq \beta$, which is clearly indicated in the Schmidt decomposition of $|\Psi_v\rangle$ as
\begin{equation}
  |\Psi_v \rangle = \lambda_0|c_0\rangle |p_-\rangle + \lambda_+|c_+\rangle |p_+\rangle + \lambda_-|c_-\rangle |p_0\rangle,   \label{varischmidt}
\end{equation}
where
\begin{eqnarray}
\lambda_0 & = & \lambda_+ = \sqrt{1-e^{-|\alpha-\beta|^2}\over 3}\\
\lambda_- & = & \sqrt{1+2e^{-|\alpha-\beta|^2}\over 3}\\
  |c_0\rangle & = & {1 \over \sqrt{3}}(|g g e \rangle + |g e g \rangle  +  |e g g \rangle),\\
  |c_+\rangle & = & {1 \over \sqrt{3}}(|g g e \rangle + \theta^2|g e g \rangle  + \theta |e g g \rangle),\\
  |c_-\rangle & = & {1 \over \sqrt{3}}(|g g e \rangle + \theta|g e g \rangle  + \theta^2 |e g g \rangle),\\
|p_0 \rangle & = & {1 \over \sqrt{3(1+2e^{-|\alpha-\beta|^2})}}(| \beta  \beta \alpha \rangle + | \beta \alpha  \beta\rangle  + |\alpha \beta \beta\rangle ),\\
|p_+ \rangle & = & {1 \over \sqrt{3(1-e^{-|\alpha-\beta|^2})}}(| \beta  \beta \alpha \rangle + \theta^2| \beta \alpha  \beta\rangle  + \theta |\alpha \beta \beta\rangle ),\\
|p_- \rangle & = & {1 \over \sqrt{3(1-e^{-|\alpha-\beta|^2})}}(| \beta  \beta \alpha \rangle + \theta| \beta \alpha  \beta\rangle  + \theta^2 |\alpha \beta \beta\rangle ).
\label{notes}
\end{eqnarray}
$\lambda_0, \lambda_+ \rightarrow 0$ for $\alpha \rightarrow \beta$, which corresponds to $\nu \rightarrow 0$.
Thus, the excitons and phonons are entangled whenever the electron-phonon interaction exists.
The entanglement entropy\cite{nielsen} $S_v$ is given by $S_v=-\lambda_0 \log \lambda_0-\lambda_+ \log \lambda_+-\lambda_- \log \lambda_-$ and shown in Figure \ref{ee} as a function of $|\alpha-\beta|$.
Figure \ref{ee} shows that $S$ lies between 0 and $\log 3$, and increases monotonically as $|\alpha-\beta|$ increases.

When the phonon degrees of freedom are traced out, we obtain the reduced density matrix as
\begin{equation}
  \rho_rv = \lambda_0^2 |c_0\rangle \langle c_0|+\lambda_+^2 |c_+\rangle \langle c_+|+\lambda_-^2 |c_-\rangle \langle c_-|,
\end{equation}
which shows that the measurement of $\chi$ gives $0$ and $\pm 1$ with corresponding probabilities.
In particular,  the average value of $\chi$ is $\chi = \lambda_+^2-\lambda_-^2=\exp (-|\alpha-\beta|^2)$.

\subsection{Transient dynamics of electron/phonon state}

In this paper, we calculated the photoinduced dynamics for two points chosen from the phase diagram (Figure \ref{phase}), one each from phases 1 and 2 in the vicinity of the phase boundary.

\begin{figure}
  \centering
  \includegraphics[width=15cm,pagebox=cropbox,clip]{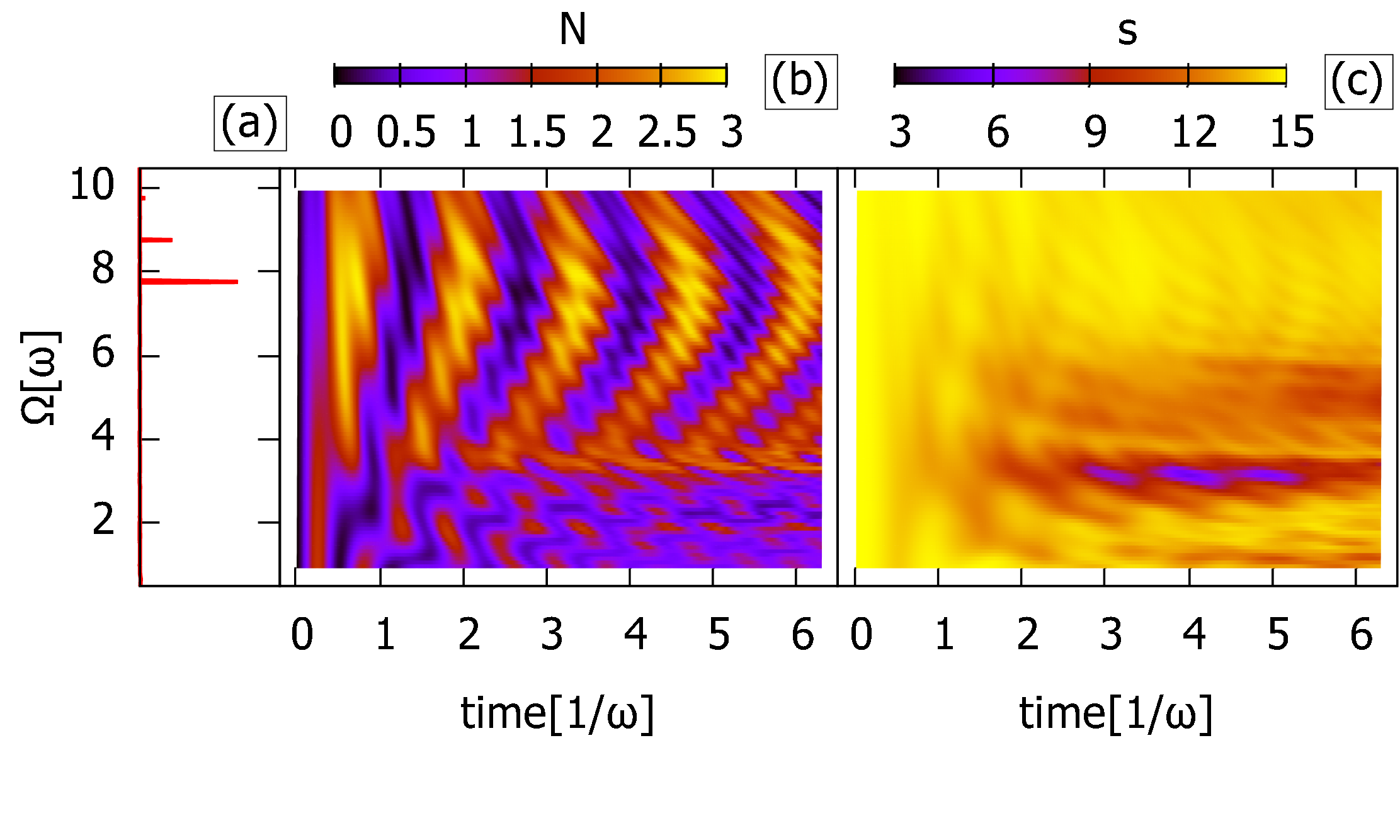}
  \caption{Calculated results for $J=1.218$, $\nu =1$, and $\mu E_0=4.75$ (phase 1). (a) absorption spectra, (b) $N$, and (c) $s$.}
  \label{pointA}
\end{figure}

Figures \ref{pointA}-(a)-(c) show the calculated results for $J = 1.218$ and $\nu = 1$, where the ordinate is the light frequency $\Omega$.
The absorption spectrum given by $A(\Omega) = \sum_n |\langle \phi_n(t)| \sum_i \sigma_x^i | \phi_n(t)\rangle|^2 \delta(\Omega-E_n)$ is represented by the bars in Figure \ref{pointA}-(a), where $E_n$ and $|\phi_n(t)\rangle$ are the $n$-th eigenvalue and the corresponding eigenvector of Hamiltonian ${\cal H}_0$, respectively.
Figures \ref{pointA}-(b) and (c) show the electronic excitation ratio $N = \langle \sum_i n_i \rangle$ and the magnitude of the total quasi-spin $\langle s \rangle$ as functions of time.
The absorption spectrum takes its maximum value at $\Omega = 7.2$, which shows that the excitation of an exciton with lattice relaxation corresponds to this peak.
We note that, the transitions to the eigenstates of the excitonic part of the Hamiltonian (\ref{ham}), i.e., $ \varepsilon \sum_j n_j + J\sum_{\langle i,j \rangle} \vec{\sigma}_i \cdot \vec{\sigma}_j$, are optically forbidden and thus do not appear in the absorption spectra.
Instead, for $\Omega > 4$, an oscillatory behavior of $N$ is observed, which is the Rabi oscillation of the exciton-phonon states mainly composed of individual excitons with lattice relaxation.
However, for $\Omega < 4$, the dynamics becomes qualitatively different and the Rabi oscillation of $N$ vanishes.
Although Figure \ref{pointA}-(a) shows that light absorption is weak for $\Omega < 8$, many energy levels corresponding to the exciton-phonon states exist as low energy excitations. 
The sudden disappearance of the Rabi oscillation indicates that such collective excitations of excitons play a major role when $\Omega < 4$,  and that the role of the dynamical behavior of exciton-phonon states should be considered.
In this case, the electronic configuration is strongly modulated by light, and the magnitude of the total quasi-spin $\langle s\rangle$ decreases at $\Omega \sim 4$ as a result. 
In other words, this feature indicates that $\nu$ mediates an effective interaction between  the states with no excitons and those with a single exciton.

Another point to note is that chirality $\chi$ remains close to zero for $1 < \Omega < 10$.
As $[{\cal H}_i, \chi] = 0$, the chirality is modulated through the electron-phonon interaction.
In phase 1, the excitons and phonons are entangled and the Schmidt decomposition of the wavefunction gives the eigenstates of $\chi$ as singular vectors.
As the singular values corresponding to $\langle \chi \rangle=1$ and $\langle \chi \rangle=-1$ are identical to each other in phase 1, the average value of the chirality is not modulated. 

\begin{figure}
  \centering
  \includegraphics[width=12cm,pagebox=cropbox,clip]{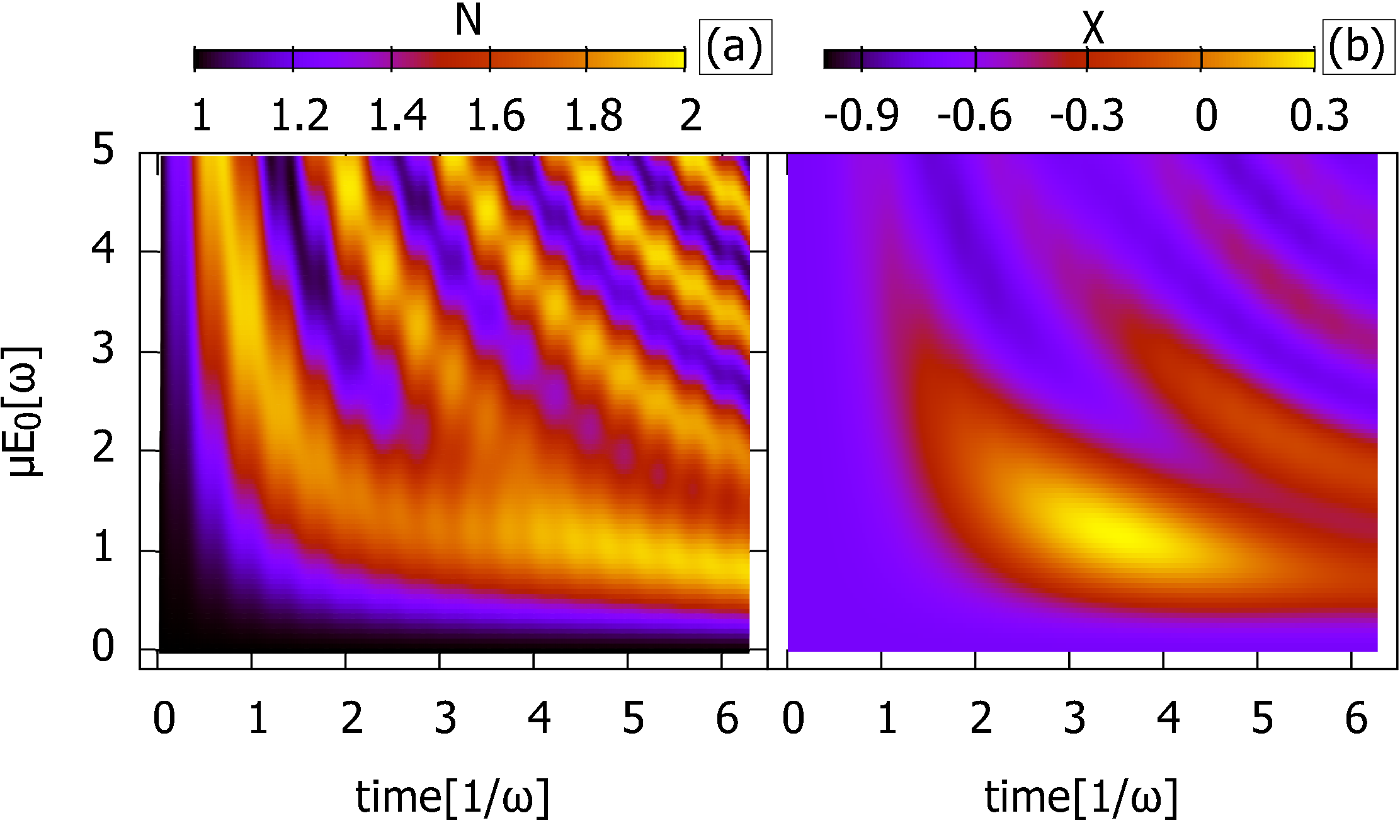}
  \caption{Calculated results for $J=1.5$, $\nu = 0.75$, and $\Omega = 8.5$ (phase 2). (a) $N$ and (b) $\langle \chi \rangle$.}
  \label{pointB}
\end{figure}
In phase 2, the exciton dynamics shows a different behavior, and to understand it, an understanding of the light intensity dependence is important.
Figure \ref{pointB}-(a) shows the value of $N$ as a function of $\mu E_0$ and time.
The frequency of light $\Omega$ is 8.5, which is chosen to maximize the modulation of $\chi$\cite{supp}.
As shown in Figure \ref{pointA}-(b), the Rabi oscillation at each site is observed in $N$ for $\mu E_0 > 2$, where the interval between peaks of $N$ increases with decreasing $\mu E_0$ because of the decrease in the Rabi frequency.
However, the Rabi oscillation of $N$ is blurred for $\mu E_0< 2$. 
A slower change of its value is observed, and it follows the lattice relaxation.
Thus, we consider that the exciton transitions follow the deformation of the lattice in this regime, and that the cooperative motion of excitons and phonons determines the dynamics of the system.
In this case, the chirality of the exciton state is modulated more strongly than $\langle s \rangle$ by light, and $\langle \chi \rangle$ changes its value as shown in Figure \ref{pointB}-(b).
More precisely, the chirality change is largest when $\mu E_0 \sim 1.8$, i.e., the  modulation of the exciton state is strongest when the Rabi oscillation vanishes.
We conclude that these results indicate that the competition between the Rabi oscillation and lattice relaxation determines exciton dynamics.

\section{Summary}

In this paper, we calculated the transient dynamics of an exciton-phonon triangular system coupled with optical phonons. 
Although $\chi$ is a good quantum number without exciton-phonon interactions, it is not conserved once electron-phonon interactions are introduced.
Therefore, we studied the excitonic properties of exciton-phonon systems with a spin-boson model to understand the control of $\chi$ by Raman scattering. 
In this case, photoirradiation causes the switching of the dipole on an ultrafast time scale, which can help us to design an “on-demand” control method of optical properties, e.g, optical rotation.
Based on the ground-state properties of the system determined by the magnitudes of the exciton-phonon interaction and the exciton transfer energy, we obtained a phase diagram in which the transition of exciton-phonon states is realized at certain points in each phase. 

The calculated results show that the frequency and/or intensity of incident light are candidates as control parameters for excitonic properties of the system, depending on the ground state character of the material, i.e., coherent control methods and the chiral properties of the ground state have a close relationship with each other.

As the phonons in the systems are not treated as reservoir, we are also interested in the coherent motion of phonons during photoexcitation.
In this case, the entanglement properties of excitons and phonons will play an important role, which is left for future study.

The material system discussed in this paper will be realized by a circular molecular aggregates\cite{mola} and/or a triple semiconductor quantum dots\cite{dots}, where the excitonic chiral properties are reflected on their optical properties. 
Since the chiral properties are discussed in this paper, we require a system without inversion symmetry.
Although the present study is not directly related with real materials, the present results contain important information on the coherent control of triangular systems. 
In particular, they will provide us a clue to design appropriate materials and control methods for chirality control.
As mentioned above, these properties will be observed in the response to the different polarization of light, for example, when a pump-probe experiment is performed.

The quantum nature of light is important in two-site systems\cite{mejpsj,mefd}, and it has been pointed out that the composite states of excitons, phonons, and photons will reveal a new feature of the coherent dynamics of materials\cite{schain,farag}.
Its role in the chiral properties of the present system will be discussed in the future study.

\medskip
\textbf{Conflict of Interest} \par
There is no conflict of interest associated with this manuscript.

\medskip
\textbf{Acknowledgements} \par 
This work was supported by JSPS KAKENHI Grant Number JP21H01752 and the Collaborative Research Project of Laboratory for Materials and Structures, Institute of Innovative Research, Tokyo Institute of Technology, Japan.

\end{document}